\documentclass[final,10pt,twocolumn]{elsarticle}
\usepackage{graphicx}
\usepackage{multirow}

\usepackage{amssymb}

\journal{Physics Letters B}

\begin{document}

\begin{frontmatter}



\title{A comparative measurement of $\phi\rightarrow K^+K^-$  
and $\phi\rightarrow \mu^+\mu^-$ in In-In collisions at the CERN SPS}


\author[INFNTO]{NA60 Collaboration\\R.~Arnaldi}
\address[INFNTO]{INFN sezione di Torino,~Italy}
\author[CERN,Heidelberg]{K.~Banicz}
\address[CERN]{CERN, 1211 Geneva 23, Switzerland}
\address[Heidelberg]{Physikalisches~Institut~der~Universit\"{a}t Heidelberg,~Germany}
\author[Clermont]{J.~Castor}
\address[Clermont]{LPC, Universit\'e Blaise Pascal and CNRS-IN2P3, Clermont-Ferrand, France}
\author[Palaiseau]{B.~Chaurand}
\address[Palaiseau]{LLR, Ecole Polytechnique and CNRS-IN2P3, Palaiseau, France}
\author[BNL]{W.~Chen}
\address[BNL]{BNL, Upton, NY, USA;}
\author[INFNCA]{C.~Cical\`o}
\address[INFNCA]{INFN sezione di Cagliari, Italy}
\author[Torino]{A.~Colla}
\address[Torino]{Universit\`a di Torino and INFN,~Italy}
\author[Torino]{P.~Cortese}
\author[CERN,Heidelberg]{S.~Damjanovic}
\author[CERN,Lisbon]{A.~David}
\address[Lisbon]{Instituto Superior T\'ecnico, Lisbon, Portugal}
\author[Cagliari]{A.~de~Falco\corref{cor1}}
\address[Cagliari]{Universit\`a di Cagliari and INFN, Cagliari, Italy}
\cortext[cor1]{Corresponding author at: Universit\`a di Cagliari and INFN, Cagliari, Italy}
\ead{alessandro.de.falco@ca.infn.it}
\author[Clermont]{A.~Devaux}
\author[Lyon]{L.~Ducroux}
\address[Lyon]{IPN-Lyon, Universit\'e Claude Bernard Lyon-I and CNRS-IN2P3, Lyon, France}
\author[RIKEN]{H.~En'yo}
\address[RIKEN]{RIKEN, Wako, Saitama, Japan}
\author[Clermont]{J.~Fargeix}
\author[Torino]{A.~Ferretti}
\author[Cagliari]{M.~Floris}
\author[CERN]{A.~F\"orster}
\author[Clermont]{P.~Force}
\author[CERN,Clermont]{N.~Guettet}
\author[Lyon]{A.~Guichard}
\author[Yerevan]{H.~Gulkanian}
\address[Yerevan]{YerPhI, Yerevan Physics Institute, Yerevan, Armenia}
\author[RIKEN]{J.~M.~Heuser}
\author[CERN,Lisbon]{M.~Keil}
\author[BNL]{Z.Li}
\author[CERN]{C.~Louren\c{c}o}
\author[Lisbon]{J.~Lozano}
\author[Clermont]{F.~Manso}
\author[CERN,Lisbon]{P.~Martins}
\author[INFNCA]{A.~Masoni}
\author[Lisbon]{A.~Neves}
\author[RIKEN]{H.~Ohnishi}
\author[INFNTO]{C.~Oppedisano}
\author[CERN,Lisbon]{P.~Parracho}
\author[Lyon]{P.~Pillot}
\author[Yerevan]{T.~Poghosyan}
\author[Cagliari]{G.~Puddu}
\author[CERN]{E.~Radermacher}
\author[CERN]{P.~Ramalhete}
\author[CERN]{P.~Rosinsky}
\author[INFNTO]{E.~Scomparin}
\author[Lisbon]{J.~Seixas}
\author[Cagliari]{S.~Serci}
\author[CERN,Lisbon]{R.~Shahoyan}
\author[Lisbon]{P.~Sonderegger}
\author[Heidelberg]{H.~J.~Specht}
\author[Lyon]{R.~Tieulent}
\author[Cagliari]{A.~Uras}
\author[Cagliari]{G.~Usai}
\author[CERN]{R.~Veenhof}
\author[Cagliari,Lisbon]{H.~K.~W\"ohri}
\date{\today}
\begin{abstract}
The NA60 experiment at the CERN SPS has studied $\phi$ meson production
in In-In collisions at 158A~GeV via both the $K^+K^-$ and the $\mu^+\mu^-$
decay channels. The yields and inverse slope parameters of the $m_T$
spectra observed in the two channels are compatible within errors,
different from the large discrepancies seen in Pb-Pb collisions 
between the hadronic (NA49) and dimuon (NA50) decay channels. Possible
physics implications 
are discussed.
\end{abstract}


\end{frontmatter}

The theory of Quantum Chromo-Dynamics (QCD) predicts that matter under extreme 
conditions of temperature and energy density undergoes a phase transition from 
hadronic matter to a plasma of deconfined quarks and gluons (QGP). 
The occurrence of this phase can be studied in the laboratory by means of high energy 
heavy-ion collisions. Strangeness enhancement has been suggested as a signature of QGP 
formation~\cite{Rafelski:1982pu}. The $\phi$ meson, being composed of an $s \overline s$ pair, 
is an ideal probe for the study of strangeness production. Moreover, the $\phi$ mass and 
branching ratios for the decay into kaon and lepton pairs may be modified in the 
medium~\cite{Lissauer:1991fr,Klingl:1997tm}.
\\
$\phi$ meson production was first studied at the CERN SPS by the NA49~\cite{Friese:2002re,Alt:2004wc}
and NA50~\cite{Alessandro:2003gy,JOU08-Phi} experiments. 
NA49 detected the $\phi$ through its decay into $K^+K^-$ pairs, while NA50 studied the 
$\phi \rightarrow \mu^+\mu^-$ channel. 
The results obtained by the two experiments in Pb-Pb collisions show discrepancies both 
in the yield and in the inverse slope parameter $T_{\mathrm{eff}}$ of the $m_T$ spectra. 
The $\phi$ multiplicity in central Pb-Pb collisions measured 
by NA50  in the dimuon channel is higher by about a factor of 4 with respect to the corresponding 
NA49 measurement in the $K^+K^-$ channel. The $T_{\mathrm{eff}}$ value found by NA50 is about 220-230~MeV, 
showing a mild dependence on centrality, while NA49 measured an inverse $m_T$ slope that increases with centrality and saturates at $T_{\mathrm{eff}}\sim 300$~MeV. It has to be stressed that the NA50 acceptance
is limited to high $p_T$, while NA49 is dominated by low $p_T$. 
\\
The discrepancy between the NA49 and NA50 results, known as ``$\phi$ puzzle", triggered a 
considerable effort to explain the observed 
differences~\cite{Shuryak:1999zh,Pal:2002aw,Santini:2006cm,Johnson:1999fv}. 
It was argued that in-medium effects may affect the spectral function of the $\phi$, causing a 
modification of its mass and partial decay widths. Moreover, kaon absorption and rescattering 
in the medium can result in a loss of signal in the region of  the $\phi$ invariant mass 
in the $K^+K^-$ channel, thus reducing the observed yield. This effect would be concentrated
at low $p_T$, causing a hardening of the $p_T$ spectrum in this
channel~\cite{Johnson:1999fv,Pal:2002aw}. 
Nevertheless, according to those calculations, the yield 
in lepton pairs is expected to exceed the one in kaon pairs  
by about $50\%$, which is much lower than the observed differences.
More recently, the CERES experiment at the SPS studied $\phi$ production in Pb-Pb collisions both in the $K^+K^-$ and dielectron channels, 
finding an agreement with the NA49 results~\cite{Adamova:2005jr}. 
However, CERES' measurement of the $T_{\mathrm{eff}}$ parameter is affected 
by a large statistical error that does not allow to draw firm conclusions.\\
The NA60 experiment measured $\phi$ production in In-In collisions at the CERN SPS. The $\phi$
meson is detected through its decay both in muon and kaon pairs. 
In this paper, results on the $\phi\rightarrow K^+K^-$ channel will be 
presented and compared to the already published ones for the 
$\phi\rightarrow\mu^+\mu^-$ decay mode~\cite{phimumu} 
and to the existing SPS measurements in different systems.
\\
The detectors relevant for the present work are  
a muon spectrometer (MS) inherited from NA50 and a vertex telescope (VT).
They are fully described in~\cite{Keil:2005zq,Usai:2005zh}. 
The muon spectrometer is composed of a toroidal magnet, 8 multi-wire
proportional chambers for the tracking and a set of scintillator hodoscopes
that provides the main trigger, which selects muon pairs coming from a 
common vertex. 
It is preceded by a 12~$\lambda_I$ thick hadron absorber. 
The VT is a high-granularity,
radiation tolerant Si pixel detector, placed between the target and the 
absorber in a 2.5~T dipole field. It is used to determine the primary vertex, 
the charged particle multiplicity and to measure the momentum of the charged 
tracks. \\
The sample used for the results presented in this paper was collected with
a 158~A$\cdot$GeV In beam impinging on a $0.17~\lambda_I$ thick In target,
composed of 7 subtargets placed in vacuum. 
The acquired statistics consists of 230 million 
triggers, mainly dimuons. More than $99\%$ of the dimuon events 
consist of uncorrelated pairs coming from the decay of pions 
and kaons in muons, or non-muon tracks. 
Since the tracks used in the present analysis are reconstructed 
using the Si tracker alone, the dimuon trigger acts like a minimum 
bias trigger. The distortion of the multiplicity distribution
introduced by the dimuon trigger drops out, since the results 
are obtained normalizing the number of $\phi \rightarrow K^+K^-$ 
obtained in each multiplicity class to the number of events 
in that class, as discussed later in the text. Therefore, 
no bias is introduced by the dimuon trigger.\\
The charged tracks associated to the vertices reconstructed in the vertex
tracker are used for this analysis. 
In order to avoid events with reinteractions of 
nuclear fragments in the subsequent targets, only events with one vertex
in the target region are selected. The vertices are required to lay inside 
one of the Indium subtargets. In addition, this cut eliminates events with 
ambiguous vertices and pile-up events, and rejects about $55\%$ of the 
statistics. 
\\
The tracks are selected requiring that the reduced $\chi^2$ 
of the track fit is lower than 3. 
The collision centrality is determined through the 
measurement of the charged particle
multiplicity. The number of produced charged particles 
$N_{ch}$ is extracted from the raw charged track 
multiplicity, applying a correction that takes into account the 
detector acceptance, reconstruction efficiency and secondary 
particle production.  The number of 
participants is then obtained assuming $N_{part} \propto dN_{ch}/d\eta$. 
We find $N_{part} \approx dN_{ch}/d\eta|_{3.7}$~\cite{phimumu}, with a 
systematic error of about $10\%$ for peripheral collisions and $5\%$ for 
central collisions. \\
The analysis in the $K^+K^-$ channel is performed assuming that all the charged 
particles associated with the primary vertex are kaons, and building all the 
possible opposite sign pairs among the tracks of each event (as discussed below, 
like sign pairs are used for systematic checks and normalization purposes). 
This results in a considerable 
combinatorial background, which is reduced by a factor 10 in the $\phi$
mass range applying a cut on the 
pair's opening angle, $0.005<\theta_{KK}<0.15$~rad. 
Due to the low acceptance for the signal at low $p_T$, and to the corresponding
high background to signal ratio (about three times higher than at high $p_T$)
a cut on the pair transverse momentum $p_T>0.9~$GeV/$c$ was applied.
Furthermore, in order to exclude the borders 
of the detector acceptance, only tracks in the 
rapidity region $2.9<y<3.7$ 
are selected, where the rapidity is calculated assigning 
the kaon mass to the tracks. After applying these 
cuts, the ratio between background and signal ranges from $\sim 190$ 
to $\sim 460$, depending on centrality.
\\
The residual background is subtracted with an event mixing technique: 
events from runs taken in homogeneous conditions are grouped in 
pools according to the position of the target associated to the vertex, the 
centrality of the collision and the direction of the event plane~\cite{Adams:2004ux}. 
The event plane azimuthal angle is calculated by means of the {\sl flow vector}.  
Its components for the second harmonic are 
$Q_2^x = \sum_i w_i \cos(2\varphi_i)$ and 
$Q_2^y = \sum_i w_i \sin(2\varphi_i)$~\cite{PhysRevC.58.1671}, 
where the weight $w_i$ associated to the $i-th$ track is 
its transverse momentum and $\varphi_i$ is its azimuthal angle.
Tracks from different events belonging to the same pool are mixed.
Each event is mixed with two other events.
Mixed events are subject to the same cuts as the ones 
applied to the real data.
\\
The obtained mixed spectra are normalized asking that in the mass region 
$1.02<m<1.06$~GeV/$c^2$ the number of mixed like-sign pairs coincides with 
the corresponding one in the real data. Changing the mass range yields 
negligible differences in the results. 
Alternatively, the normalization factor is chosen such that the integral of the opposite 
sign mass spectra in the whole mass range is equal for the real and mixed samples. 
For both methods, the normalization factor is evaluated either for each subtarget 
separately, or averaged over the targets. The differences in the results coming from 
the choice of the normalization criterion, causing variations of about $4\%$ in the 
determination of the number of $\phi$ mesons per centrality bin, are taken into account 
in the systematic error.\\ 
%
%
\begin{figure}[t!]
\begin{center}
\includegraphics*[width=0.42\textwidth]{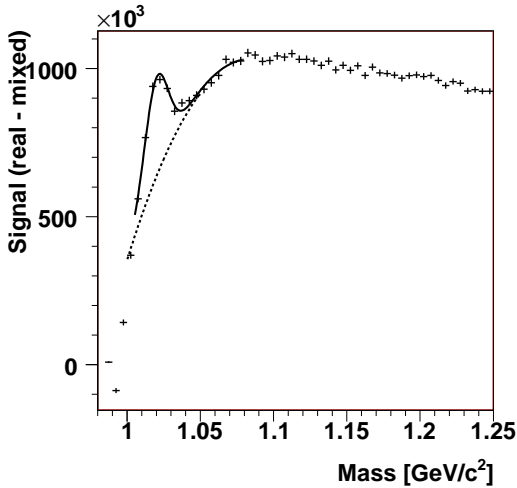}
\caption{Invariant mass spectrum of the opposite sign pairs after 
combinatorial background subtraction for $p_T>0.9$~GeV/$c$ integrated 
in centrality.}
\label{fig:massSpectrum}
\end{center}
\end{figure}
%
%
%
%
\begin{table*}
\caption{Centrality bins, measured number of $\phi$, ratio between background and signal, measured 
$\phi$ mass and width.}
\centering
\begin{tabular}{cccccc}
\hline 
$dN_{ch}/d\eta$ range &  $\langle N_{part} \rangle$ &  $N_\phi$(meas.) & B/S  & $m_\phi$ (MeV/$c^2$) &
 $\sigma_m$ (MeV/$c^2$) \\
\hline  
29-42      &    41       & $1.5 \cdot 10^4$ & 188 & \multirow{2}{*}    {$1019.9 \pm 1.7$}  &   \multirow{2}{*}{$6.5 \pm 1.2 $}\\
42-95      &    78       & $1.3 \cdot 10^5$ & 240 &  & \\
96-160    &   133      & $6.5 \cdot 10^5$ & 310                                   & $1019.9 \pm 1.2$        & $7.5\pm 1.1$\\
161-250  &   177      & $6.8 \cdot 10^5$ & 465                                    & $1019.3\pm 1.8$         & $7.4\pm 2.3$\\

\hline 
\end{tabular}
\label{table:centrality}
\vspace{-0.3cm}
\end{table*}
%
%
The invariant mass spectrum of the opposite sign pairs after combinatorial background 
subtraction (normalized with the like-sign pairs) is shown in Fig.~\ref{fig:massSpectrum}, 
integrated in centrality. 
%
The function for the $\phi$ peak is determined fitting the mass distribution obtained 
with an overlay Monte Carlo simulation, that consists in reconstructing a generated 
$\phi$ decaying into a kaon pair on top of a real event. We use a gaussian 
superimposed to an empirical function that takes into account the 
mass tails and accounts for about $5\%$ of the total number of $\phi$. 
%
Several functions and fit mass ranges have been tested to describe 
the residual background, including polynomials of first and second order and
functions that are null at the $KK$ mass threshold. 
In order to check that the fit functions correctly reproduce the signal and the
residual background, the fit procedure has been applied to the like-sign 
invariant mass spectra in several $p_T$ intervals, after the subtraction 
of the event-mixed spectra. The like-sign pairs spectra have 
a shape which is very similar to the corresponding opposite 
sign ones for masses above the $\phi$ peak ($m>1.04$~GeV/$c^2$), 
and show a smooth trend down to $m=1$~GeV/$c^2$. 
Since these pairs do not contain any signal, then a fit applied to them should
result in a null signal component, while the presence of a fictitious peak would 
indicate a bias in the fitting procedure. 
The fit applied to the like-sign pairs gives a signal component compatible with 
zero in all the $p_T$ bins, showing that no artifact is introduced by the choice 
of the fit function.\\
It has to be noted that if the $\phi$ peak position and width are left 
as free parameters of the fit, the corresponding
values are $m_\phi = 1019.5 \pm 0.3 \pm 1.2$~MeV/$c^2$ and 
$\sigma_m=7.8 \pm 0.3 \pm 1.2$~MeV/$c^2$, in good agreement with the simulations.
Corresponding results as a function of centrality are reported in 
Table~\ref{table:centrality}, together with the number 
of reconstructed $\phi$ mesons, obtained by integrating the 
function describing the $\phi$ peak, and the ratio between background 
and signal in the mass range $1.005<m<1.035$~GeV/$c^2$. Statistical 
and systematic errors for $m_\phi$ and $\sigma_m$ are added in quadrature.
The values show no dependence on 
centrality, indicating that no modification of  the spectral function is visible 
within the sensitivity of the measurement. A similar result is obtained in the 
dimuon channel~\cite{phimumu}. 
The results shown in the following are obtained fixing $m_\phi$ and $\sigma_m$ 
to 1.019~GeV/$c^2$ and 7.8~MeV/$c^2$, according to the Monte Carlo values. \\
%
%
\begin{figure}[t!]
\begin{center}
\includegraphics*[width=0.42\textwidth]{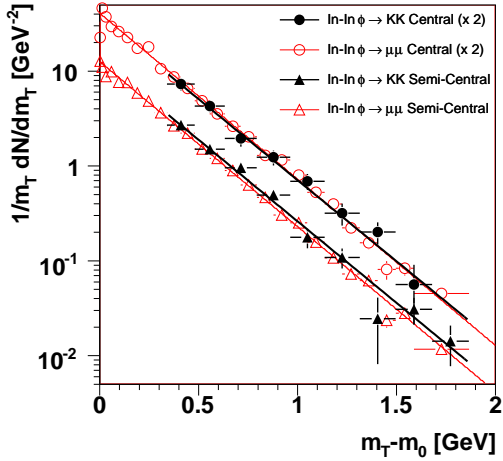}
\caption{Normalized transverse mass spectra for semi-central (triangles) and central (circles)
collisions for the $\phi\rightarrow K^+K^-$ channel (full symbols) compared to the 
corresponding ones in muon pairs (open symbols)~\cite{phimumu}. 
Results for central collisions are scaled by a factor 2 in order to improve readibility.
Colour online.}
\label{fig:mtspectra}
\end{center}
\end{figure}
%
%
%
%
\begin{table*}
\caption{$\phi$ inverse slope and yield extracted from the analyses in the muon and kaon channels 
as a function of centrality.}
\centering
\begin{tabular}{cccccc}
\hline 
\small $\langle N_{part} \rangle$ & $T^{\mu\mu}_{\mathrm{eff}}$  (MeV) &    $\langle \phi \rangle_{\mu\mu}(p_T>0.9$~GeV$)$ & $T^{KK}_{\mathrm{eff}}$  (MeV) &    $\langle \phi \rangle_{KK}(p_T>0.9$~GeV$)$ &  $\langle \phi \rangle_{KK}(p_T>0)$\\
\hline  
15   & $209 \pm 4$  & $0.044 \pm 0.002 \pm 0.005$ &     -              & - & - \\  
41   & $232 \pm 4$  & $0.197 \pm 0.009 \pm 0.021$ &     -              & $0.16 \pm 0.05 \pm 0.08$  & $0.64 \pm 0.17 \pm 0.03$ \\
78   & $245 \pm 4$  & $0.48  \pm 0.02  \pm 0.05$  &     -              & $0.47 \pm 0.06 \pm 0.02$ & $1.60 \pm 0.18 \pm 0.06$\\
133  & $250 \pm 4$  & $1.03  \pm 0.03  \pm 0.07$  & $253 \pm 11 \pm 5$ & $1.01 \pm 0.07 \pm 0.05 $ &$3.5 \pm 0.2 \pm 0.2$\\
177  & $249 \pm 5$  & $1.65  \pm 0.06  \pm 0.07$  & $254 \pm 13 \pm 6$ & $1.49 \pm 0.11 \pm 0.07$ &$4.6 \pm 0.3 \pm 0.4$\\
\hline 
\end{tabular}
\label{table:results}
\vspace{-0.3cm}
\end{table*}
%
%
In order to extract the $m_T$ distributions, the fit to the invariant mass 
spectra is performed in several $p_T$ intervals having a size of 200~MeV/$c$.  
It was checked that results do not depend on the choice of the bin size. 
The $m_T$ distributions are then corrected for the geometrical acceptance 
and reconstruction efficiency with an overlay 
Monte Carlo simulation.
The rapidity and decay angle distributions used 
for the $\phi$ generation are tuned to the ones measured in muons. 
The former is a gaussian with $\sigma_y=1.13$, while the $dN/d\cos{\theta}$ 
distribution is assumed to be flat~\cite{phimumu}.

The acceptance in multiplied by reconstruction efficiency 
is almost flat for $p_T>0.9$~GeV/$c$, about $15\%$ in
full rapidity. 
A reliable $m_T$ distribution could be extracted only for 
semicentral ($\langle N_{part}\rangle=133$) and central ($\langle N_{part}\rangle=177$) 
collisions, while for semiperipheral collisions ($\langle N_{part}\rangle=78$), due 
to statistics limitations, the resulting inverse slope parameter was not stable 
when varying the analysis criteria. 
Results are reported in 
Fig.~\ref{fig:mtspectra}, where the distributions in the 
$\phi\rightarrow \mu \mu$ channel in the same 
centrality intervals are reported for comparison. All the spectra are normalized 
to the $\phi$ multiplicity as shown in Table ~\ref{table:results} (details on the method 
used to measure the $\phi$ yield 
are reported below). 
The $m_T$ spectra are fitted with the function
$1/m_T dN/dm_T \propto e^{-m_T/T_{\mathrm{eff}}}$. 
The reduced $\chi^2$ is $\sim 1$ in both cases.
Results are reported in Table~\ref{table:results}.
The systematic error
is dominated by the choice of the function used for the background. 
The systematic errors due to the choice of the normalization criterion, 
to the variation of the analysis cuts and of the starting point of the 
fit to the mass spectra are also taken into account.  
Results are in good agreement with the ones obtained in dimuons in the 
full $p_T$ range, also reported 
in Table~\ref{table:results}. Since in 
presence of radial flow the $T_{\mathrm{eff}}$ value may depend on the $p_T$ range,
the fit to the dimuon spectra was 
restricted to the range $p_T>0.9$~GeV/$c$, giving $T_{\mathrm{eff}}=252\pm 4$ 
and $247 \pm 3$~MeV for semicentral and central collisions, still in agreement
with the measurement in kaons.\\ 
%
%
\begin{figure}[t!]
\begin{center}
\includegraphics*[width=0.42\textwidth]{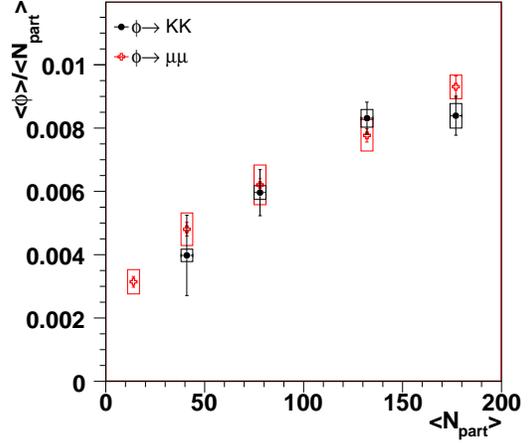}
\caption{$\langle \phi \rangle /N_{part}$
as a function of the number of participants in In-In collisions in the 
$\phi\rightarrow K^+K^-$ (full circles) and $\phi\rightarrow \mu\mu$ (open crosses) 
channels for $p_T>0.9$~GeV$/c$. {\sl Boxes:} systematic errors. Colour online.}
\label{fig:yield}
\end{center}
\end{figure}
%
%
The raw $\phi$ multiplicity is determined fitting the mass spectra for 
$p_T>0.9$~GeV/$c$ and dividing the number of $\phi$ mesons obtained by the 
total number of events selected for this analysis. This value is then corrected 
for the branching ratio in kaon pairs, $(49.2 \pm 0.6)\%$~\cite{Yao:2006px} and 
for the acceptance, evaluated through a Monte Carlo simulation. 
The inverse $m_T$ slopes used for the calculation of the acceptance are the 
ones measured in the kaon channel, where available; otherwise the values 
obtained in dimuons are used. 
Table~\ref{table:results} reports the results for the $\phi$ multiplicity in 
the dimuon and kaon channels for $p_T>0.9$~GeV$/c$.   
The main contributions to the systematic error are given by 
the uncertainty in the choice of the fit function 
for the residual background component in the fit of the invariant mass spectra
and the normalization criterion. 
The systematic errors due to the variation of the analysis cuts 
and of the starting point of the fit to the mass spectra are also 
taken into account.\\
In Fig.~\ref{fig:yield} the ratio $\langle \phi \rangle /N_{part}$ 
as a function of $N_{part}$ in the $\phi\rightarrow K^+K^-$ channel 
for $p_T>0.9$~GeV$/c$ is compared to the corresponding one in dimuons. 
The additional contribution due to the systematic error in $N_{part}$, 
affecting both channels in the same way, is not displayed. 
It can be seen that the yield in the hadronic channel  
and the one in the dileptonic channel are in agreement within the errors.
A ratio between the $\phi$ yields in dimuons and in kaons 
larger than 1.18 in the common $p_T$ range is excluded at $95\%$ C.L.\\
The $\phi$ multiplicity in kaons was also calculated in the  
full $p_T$ range. Results are reported in the last column of 
Table~\ref{table:results}.
The systematic error due to the uncertainty in the extrapolation to 
full $p_T$ caused by the error in the inverse slope was taken into
account. 
It has to be stressed that the extrapolation to $p_T=0$ was done 
under the hypothesis that the inverse slope does not change. 
Models like AMPT~\cite{Pal:2002aw} predict that kaon rescattering 
causes a depletion in the number of reconstructed $\phi\rightarrow KK$
which is concentrated at low $p_T$, causing an increase of $T_{\mathrm{eff}}$
in that region. 
In Pb-Pb central collisions, the effect would be already visible for 
$m_T-m_0 \geq 0.34$~GeV, corresponding to the region covered in this 
analysis. In that region, the difference between $T_{\mathrm{eff}}$ in 
kaon and muon pairs would range from 30 to 50 MeV, and the 
fractional loss in kaons at $m_T-m_0 \sim 0.34$~GeV would range 
from $35\%$ to $50\%$. Such effects are not seen in our data, 
suggesting that if any rescattering effect is present, it 
would be concentrated at lower $p_T$. In the range 
$0<m_T-m_0 < 0.34$~GeV the AMPT model predicts an inverse 
$m_T$ slope of about 330~MeV, close to the value observed by 
NA49 in central Pb-Pb collisions. Since no theoretical 
predictions for In-In collisions are available, we estimated 
the effect of the change of slope in the extrapolation to 
$p_T=0$ under the extreme hypothesis $T_{\mathrm{eff}}=330$~MeV
for $0<m_T-m_0 < 0.34$~GeV, while for $m_T-m_0 \geq 0.34$~GeV
the measured value is kept. This variation would cause a 
reduction of the $\phi$ yield in kaons of about $12\%$. 
This quantity can be considered as a conservative estimation 
of the uncertainty in the extrapolation due to a possible 
suppression mechanism of the hadronic channel.\\
%
%
 %
%
\begin{figure}[t!]
\begin{center}
\includegraphics*[width=0.49\textwidth]{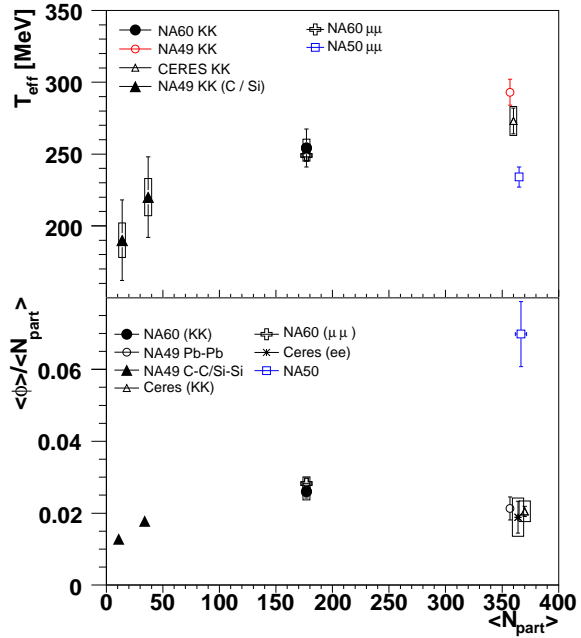}
\caption{Inverse slope (top) and $\langle \phi \rangle /N_{part}$ 
in full $p_T$ range (bottom) as a function of $N_{part}$ 
for central collisions. 
{\sl Boxes:} systematic errors. Colour online.}
\label{fig:collSyst}
\end{center}
\end{figure}
%
%
In order to compare to other collision systems, the inverse slope and
the enhancement, quantified as the ratio $\langle \phi \rangle /
N_{part}$ in the full $p_T$ range, are plotted in 
Fig.~\ref{fig:collSyst} as a function of the
number of participants for central C-C, Si-Si, In-In and Pb-Pb
collisions~\cite{Friese:2002re,Alt:2004wc}. For this comparison, 
the systematic error on $N_{part}$,
of about $5\%$ in central collisions, is taken into account in the 
calculation of the ratio $\langle \phi \rangle /N_{part}$ for the 
NA60 points. 
\\ 
The inverse slope shows an initial fast increase at low
$N_{part}$ values, that becomes less pronounced going towards higher
$N_{part}$.
A lower value is observed by NA50 as compared both to the CERES and
NA49 measurements in Pb-Pb and to the NA60 In-In points in the
hadronic and dileptonic channels.

As stated above, in the presence of radial flow $T_{\mathrm{eff}}$
depends on $p_T$.  The NA60 analysis in dimuons~\cite{phimumu},
performed at low and high $p_T$ ranges, shows a difference limited to
about 15~MeV in In-In. 
This fact, complemented by the detailed blast wave analysis of 
the dimuon data discussed in~\cite{phimumu}, makes it difficult 
to ascribe only to radial flow the large differences in  
$T_{\mathrm{eff}}$ shown in Fig.~\ref{fig:collSyst}.
A further flattening caused by kaon rescattering and
absorption may lead to larger $T_{\mathrm{eff}}$ values in the
hadronic channel in Pb-Pb. \\
Concerning the enhancement, the NA60 measurements in the dilepton 
and hadron channels can differ up to $18\%$ (at $95\%$ CL) 
for $p_T>0.9$~GeV/$c$ and $22\%$ in full $p_T$ considering 
the uncertainty in the extrapolation arising from
a possible suppression of the hadronic channel at low $p_T$.
In Fig.~\ref{fig:collSyst} the NA50 result is extrapolated to full phase
space using $T_{\mathrm{eff}}=220$~MeV, according to the NA50 measurement in
peripheral collisions.  Even assuming as an extreme case
$T_{\mathrm{eff}}=300$~MeV, as obtained by NA49 in central collisions, the NA50
enhancement would exceed by a factor of $\sim 2$ the central Pb-Pb
values measured in kaons.  
\\ 
The yield per participant measured in In-In exceeds the one observed 
in Pb-Pb in the
hadronic channel (both NA49 and CERES) by about 30$\%$. This might
suggest a suppressing mechanism for the kaon channel, below
experimental sensitivity in In-In, which shows up in Pb-Pb.  The CERES
measurement in dielectrons is in agreement with the one in kaons but,
given the measurement errors, it cannot rule out differences of the order
of 40-50\%, expected by models including kaon rescattering. If one
assumes that no suppression mechanism affects the Pb-Pb collisions, it is
then difficult to understand why the $\phi$ multiplicity in central
Pb-Pb is smaller than in central In-In.  
\\ 
In conclusion, the inverse slopes and yields measured in the kaon and
muon channels in In-In collisions are in agreement, excluding for central
collisions a difference in the yields larger than $18~\%$ 
(at $95~\%$ C.L.) in the common $p_T$ range. 
In addition, no modification of the $\phi$ mass and width is observed as
a function of centrality.  When the comparison is extended to other
systems, it is difficult to reconcile all of the observations into a
coherent picture, albeit there is some hint for a possible physics
mechanism leading to a difference in the two channels in Pb-Pb
collisions, while producing no remarkable difference in In-In
collisions. 
\\
%
%
%
%
%
\bibliographystyle{model1a-num-names}
\bibliography{phikkPhysLett}

\begin{thebibliography}{18}
\expandafter\ifx\csname natexlab\endcsname\relax\def\natexlab#1{#1}\fi
\providecommand{\bibinfo}[2]{#2}
\ifx\xfnm\relax \def\xfnm[#1]{\unskip,\space#1}\fi
\bibitem[{Rafelski and Muller(1982)}]{Rafelski:1982pu}
\bibinfo{author}{J.~Rafelski}, \bibinfo{author}{B.~Muller},
  \bibinfo{journal}{Phys. Rev. Lett.} \bibinfo{volume}{48}
  (\bibinfo{year}{1982}) \bibinfo{pages}{1066}.
\bibitem[{Lissauer and Shuryak(1991)}]{Lissauer:1991fr}
\bibinfo{author}{D.~Lissauer}, \bibinfo{author}{E.~V. Shuryak},
  \bibinfo{journal}{Phys. Lett.} \bibinfo{volume}{B253} (\bibinfo{year}{1991})
  \bibinfo{pages}{15--18}.
\bibitem[{Klingl et~al.(1998)Klingl, Waas, and Weise}]{Klingl:1997tm}
\bibinfo{author}{F.~Klingl}, \bibinfo{author}{T.~Waas},
  \bibinfo{author}{W.~Weise}, \bibinfo{journal}{Phys. Lett.}
  \bibinfo{volume}{B431} (\bibinfo{year}{1998}) \bibinfo{pages}{254--262}.
\bibitem[{Friese(2002)}]{Friese:2002re}
\bibinfo{author}{V.~Friese}, \bibinfo{journal}{Nucl. Phys.}
  \bibinfo{volume}{A698} (\bibinfo{year}{2002}) \bibinfo{pages}{487--490}.
\bibitem[{Alt et~al.(2005)}]{Alt:2004wc}
\bibinfo{author}{C.~Alt}, et~al., \bibinfo{journal}{Phys. Rev. Lett.}
  \bibinfo{volume}{94} (\bibinfo{year}{2005}) \bibinfo{pages}{052301}.
\bibitem[{Alessandro et~al.(2003)}]{Alessandro:2003gy}
\bibinfo{author}{B.~Alessandro}, et~al., \bibinfo{journal}{Phys. Lett.}
  \bibinfo{volume}{B555} (\bibinfo{year}{2003}) \bibinfo{pages}{147--155}.
\bibitem[{Jouan et~al.(2008)}]{JOU08-Phi}
\bibinfo{author}{D.~Jouan}, et~al., \bibinfo{journal}{J. Phys.s}
  \bibinfo{volume}{G35} (\bibinfo{year}{2008}) \bibinfo{pages}{104163+}.
\bibitem[{Shuryak(1999)}]{Shuryak:1999zh}
\bibinfo{author}{E.~V. Shuryak}, \bibinfo{journal}{Nucl. Phys.}
  \bibinfo{volume}{A661} (\bibinfo{year}{1999}) \bibinfo{pages}{119--129}.
\bibitem[{Pal et~al.(2002)Pal, Ko, and Lin}]{Pal:2002aw}
\bibinfo{author}{S.~Pal}, \bibinfo{author}{C.~M. Ko}, \bibinfo{author}{Z.-w.
  Lin}, \bibinfo{journal}{Nucl. Phys.} \bibinfo{volume}{A707}
  (\bibinfo{year}{2002}) \bibinfo{pages}{525--539}.
\bibitem[{Santini et~al.(2006)Santini, Burau, Faessler, and
  Fuchs}]{Santini:2006cm}
\bibinfo{author}{E.~Santini}, \bibinfo{author}{G.~Burau},
  \bibinfo{author}{A.~Faessler}, \bibinfo{author}{C.~Fuchs},
  \bibinfo{journal}{Eur. Phys. J.} \bibinfo{volume}{A28} (\bibinfo{year}{2006})
  \bibinfo{pages}{187--192}.
\bibitem[{Johnson et~al.(2001)Johnson, Jacak, and Drees}]{Johnson:1999fv}
\bibinfo{author}{S.~C. Johnson}, \bibinfo{author}{B.~V. Jacak},
  \bibinfo{author}{A.~Drees}, \bibinfo{journal}{Eur. Phys. J.}
  \bibinfo{volume}{C18} (\bibinfo{year}{2001}) \bibinfo{pages}{645--649}.
\bibitem[{Adamova et~al.(2006)}]{Adamova:2005jr}
\bibinfo{author}{D.~Adamova}, et~al., \bibinfo{journal}{Phys. Rev. Lett.}
  \bibinfo{volume}{96} (\bibinfo{year}{2006}) \bibinfo{pages}{152301}.
\bibitem[{Arnaldi et~al.(2009)}]{phimumu}
\bibinfo{author}{R.~Arnaldi}, et~al., \bibinfo{journal}{Eur. Phys. J.}
  \bibinfo{volume}{C64} (\bibinfo{year}{2009}) \bibinfo{pages}{1--18}.
\bibitem[{Keil et~al.(2005)}]{Keil:2005zq}
\bibinfo{author}{M.~Keil}, et~al., \bibinfo{journal}{Nucl. Instrum. Meth.}
  \bibinfo{volume}{A546} (\bibinfo{year}{2005}) \bibinfo{pages}{448--456}.
\bibitem[{Usai et~al.(2005)}]{Usai:2005zh}
\bibinfo{author}{G.~Usai}, et~al., \bibinfo{journal}{Eur. Phys. J.}
  \bibinfo{volume}{C43} (\bibinfo{year}{2005}) \bibinfo{pages}{415--420}.
\bibitem[{Adams et~al.(2005)}]{Adams:2004ux}
\bibinfo{author}{J.~Adams}, et~al., \bibinfo{journal}{Phys. Lett.}
  \bibinfo{volume}{B612} (\bibinfo{year}{2005}) \bibinfo{pages}{181--189}.
\bibitem[{Poskanzer and Voloshin(1998)}]{PhysRevC.58.1671}
\bibinfo{author}{A.~M. Poskanzer}, \bibinfo{author}{S.~A. Voloshin},
  \bibinfo{journal}{Phys. Rev. C} \bibinfo{volume}{58} (\bibinfo{year}{1998})
  \bibinfo{pages}{1671--1678}.
\bibitem[{Yao et~al.(2006)}]{Yao:2006px}
\bibinfo{author}{W.~M. Yao}, et~al., \bibinfo{journal}{J. Phys.}
  \bibinfo{volume}{G33} (\bibinfo{year}{2006}) \bibinfo{pages}{1--1232}.

\end{thebibliography}
\end{document}